# Experimental measurement of non-Hermitian left eigenvectors

Xulong Wang[1,2], Guancong Ma[1, 3†]

[1]Department of Physics, Hong Kong Baptist University, Kowloon Tong, Hong Kong, China
[2]Department of Physics, Hong Kong University of Science and Technology, Clear Water Bay, Kowloon, Hong Kong, China
[3]Shenzhen Institute for Research and Continuing Education, Hong Kong Baptist University, Shenzhen 518000, China

[†] Corresponding author E-mail: phgcma@hkbu.edu.hk

**Abstract**

The duality of left and right eigenvectors underpins the comprehensive understanding of many physical phenomena. In Hermitian systems, left and right eigenvectors are simply Hermitian-conjugate pairs. Non-Hermitian eigenstates, in contrast, have left and right eigenvectors that are distinct from each other. However, despite the tremendous interest in non-Hermitian physics in recent years, the roles of non-Hermitian left eigenvectors (LEVs) are still inadequately explored – their physical consequences and observable effects remain elusive, so much so that LEVs seem largely like an object of primarily mathematical purpose. In this study, we present a method based on the non-Hermitian Green's function for directly retrieving both LEVs and REVs from experimentally measured steady-state responses. We validate the effectiveness of this approach in two separate acoustic experiments: one characterizes the non-Hermitian Berry phase, and the other measures extended topological modes. Our results not only unambiguously demonstrate observable effects related to non-Hermitian LEVs, but also highlight the under-appreciated role of LEVs in non-Hermitian phenomena.

## 1. Introduction

The study of non-Hermitian systems has gained tremendous attention in recent years [1–6]. Non-Hermitian operators are fundamentally distinct from Hermitian ones because their eigenvalues are generically complex, and their eigenvectors are nonorthogonal [7–9]. The relations between Hermitian and non-Hermitian eigenstates are shown in Fig. 1(a). The complex eigenvalues are the root of many intriguing phenomena, such as those relating to spectral branch-point singularities known as exceptional points (EPs) [5,9]. Equally important is the nonorthogonality of eigenvectors, which underlies state-switching behavior when an EP is encircled [10–14] and gives rise to the fractional Berry phases [5,15]. It is also what fundamentally permits the existence of non-Hermitian skin effect (NHSE) [16–18] – indeed, if the eigenvectors were required to be mutually orthogonal, i.e., each having zero overlap with another one, it is impossible for all of them to exponentially congregate at the boundary. The accurate description of non-Hermitian systems requires both the left eigenvectors (LEVs) and right eigenvectors (REVs). Many non-Hermitian physical laws and phenomena hinge on the joint consideration of LEVs and REVs. For example, they together define the bi-orthonormal condition, which underpins entities such as the bi-orthogonal inverse participation ratio that measures the degree of localization of non-Hermitian states [19,20], bi-orthogonal bulk-boundary correspondence that predicts topological transition in the presence of NHSE [21], and non-Hermitian geometric phase that characterizes the topology of EPs [22]. However, unlike REVs that correspond to wavefunctions that are directly observable (at least in classical settings) [23–27], LEVs have mostly been considered for mathematical reasons, e.g., bi-orthogonal normalization. The physical implications or observable effects of LEVs are so far still elusive. To understand why this is the case, we begin with the definition of REVs $|\psi^R\rangle$ and LEVs $\langle\psi^L|$,

$$\widehat{H}|\psi_n^R\rangle = \omega_n|\psi_n^R\rangle, \qquad \langle\psi_n^L|\widehat{H} = \omega_n\langle\psi_n^L|, \tag{1}$$

where $\widehat{H}$ is a non-Hermitian Hamiltonian, $\omega_n$ is the eigenvalue of the *n*-th eigenstate, and $|\cdot\rangle = \langle\cdot|^\dagger$. LEVs and REVs are not Hermitian-conjugates of one another, i.e., $|\psi_n^R\rangle \neq \langle\psi_n^L|^\dagger$. This can be seen from the fact that $\widehat{H} \neq \widehat{H}^\dagger$ and $\omega_n \neq \omega_n^*$, so $\widehat{H}^\dagger|\psi_n^L\rangle = \omega_n^*|\psi_n^L\rangle$, i.e., $|\psi_n^L\rangle$ is an REV of a different system described by $H^\dagger$. The relation and difference of REVs and LEVs are graphically shown in Figs. 1(a, b). It is worth mentioning that, in the presence of certain symmetries, LEVs can be obtained from REVs. For example, if the Hamiltonian is symmetric, e.g., for reciprocal non-Hermitian systems, then $\widehat{H}(k) = \widehat{H}^T(-k)$, such that $\langle\psi_n^L| = |\psi_n^R\rangle^T$ [15,28]. However, convenient conversions like this do not generically exist.

Therefore, one needs to separately construct a different system $\widehat{H}^\dagger$, then deduce from it the LEV of $\widehat{H}$. This is often unrealistic in experiments. For these fundamental and realistic reasons, non-Hermitian LEVs have so far been overlooked in experiments and remain primarily a theoretical construct.

In this work, we present the first experimental measurement of LEVs in non-Hermitian systems. Our approach is based on the universal appearance of LEVs in non-Hermitian Green's functions [9,29], through which the LEVs can be retrieved from the measured steady-state responses between multiple excitation sources at different locations and a single receiver. Using this approach, we successfully obtained the LEVs in two different acoustic experiments: one is a two-level non-Hermitian system without any symmetry, and the other is a one-dimensional (1D) non-Hermitian Su-Schrieffer-Heeger (SSH) lattice sustaining an extended topological mode (as a REV). Our work completes the full picture of non-Hermitian physics and underscores the importance of LEVs for comprehensively understanding non-Hermitian physics.

## 2. Eigenvectors and non-Hermitian Green's function.

We begin with the Green's function of an $N$-level system

$$\widehat{G}(\omega) = \frac{1}{\omega - \widehat{H}} = \sum_{n=1}^{N} \frac{A_0 |\psi_n^R\rangle\langle\psi_n^L|}{\omega - \omega_n}. \tag{2}$$

where $\omega \in \mathbb{R}$ is the excitation angular frequency, $A_0$ is a constant that accounts for the properties of sources and probes. The steady-state response is obtained as $A(\omega) = \langle P|\widehat{G}(\omega)|S\rangle$, where $|P\rangle$ and $|S\rangle$ are vectors representing the probe and the source, respectively. For convenience and without loss of generality, $|P\rangle$ and $|S\rangle$ are $N \times 1$ in dimension. Let us consider the consequence of interchanging $|P\rangle$ and $|S\rangle$. If $\widehat{H}$ is Hermitian, then $|\psi_n^R\rangle = \langle\psi_n^L|^\dagger$ so $\widehat{G}(\omega)$ is a symmetric operator. In other words, switching $|P\rangle$ and $|S\rangle$ does not affect the response $A(\omega)$. If $\widehat{H}$ is non-Hermitian and asymmetric, e.g., contains non-reciprocity in hopping terms, $\widehat{G}(\omega)$ is also asymmetric. As a result, switching $|P\rangle$ and $|S\rangle$ yields different $A(\omega)$. In fact, Eq. (2) encodes REVs (LEVs) in the columns (rows) of $\widehat{G}(\omega)$, such that entries in REVs can be extracted by fixing the position of the source and then varying the probe, whereas LEVs can be retrieved by fixing the probe and changing the position of the source. These are illustrated in Figs. 1(c, d).

Thus, we arrive at a procedure for obtaining the LEVs from response measurements. We illustrate using a simple two-site non-Hermitian system, whose REVs are $|\psi_n^R\rangle = (r_n^1 \quad r_n^2)^T$, and LEVs are $\langle\psi_n^L| = (l_n^1 \quad l_n^2)$. (We neglect the normalization factors here for brevity.) For ease of data processing, only one probe and one source are used, so there is only one non-zero entry in $|P_j\rangle$ and $|S_i\rangle$, which is indicated by the subscript. The measurement then has two steps. In the first step, we place both the source and probe at site-1, so $|S_1\rangle = |P_1\rangle = (1 \quad 0)^T$. The measured response is $A_{11}(\omega) = \langle P_1|\hat{G}(\omega)|S_1\rangle = G_{11}(\omega) = \frac{A_0 r_1^1 l_1^1}{\omega - \omega_1} + \frac{A_0 r_2^1 l_2^1}{\omega - \omega_2}$. In the second step, the source is moved to site-2 whereas the probe is unchanged, i.e., $|S_2\rangle = (0 \quad 1)^T$, $|P_1\rangle = (1 \quad 0)^T$, which produces $A_{12}(\omega) = \langle P_1|\hat{G}(\omega)|S_2\rangle = G_{12}(\omega) = \frac{A_0 r_1^1 l_1^2}{\omega - \omega_1} + \frac{A_0 r_2^1 l_2^2}{\omega - \omega_2}$. From the two measured spectra responses, pick the data point from the first resonant peak, i.e., $\omega \cong \mathrm{Re}(\omega_1)$, such that the response from the second eigenmode can be neglected, we can obtain the two entries in $\langle \psi_1^L|$, up to a common pre-factor $\frac{A_0 r_1^1}{\omega - \omega_1}$. Likewise, the entries in $\langle \psi_2^L|$ can be obtained from the second resonant peaks, with a common pre-factor $\frac{A_0 r_2^1}{\omega - \omega_2}$. (If the resonant peaks overlap, numerical fitting is required to precisely obtain the entries.) It is straightforward to see that entries of the REVs can be retrieved likewise, but the procedure changes the position of the probe and keeps the source fixed.

## 3. Experimental retrieval of the LEVs

### 3.1 LEVs of a two-level non-Hermitian system

We now experimentally apply the method delineated above to two different acoustic experiments. In the first experiment, we aim to obtain the non-Hermitian Berry phase of an order-2 EP. It is well-known that the Berry phase characterizes the parallel transport of eigenvectors as fiber bundles on a base manifold,

$$\Theta = i \int_C d\vec{\lambda}\, \langle \psi_n^L(\vec{\lambda})|\nabla_\lambda|\psi_n^R(\vec{\lambda})\rangle, \qquad (3)$$

where $\vec{\lambda}$ is the parameter, and $C$ denotes the evolution path. Apparently, both REV and LEV are needed to obtain $\Theta$.

$$\hat{H}_1(\phi_x, \phi_y) = (\omega_0 + i\gamma_0)I_2 + \begin{pmatrix} i\gamma_1 & \phi_x \\ \phi_x + \gamma_2 & \phi_y \end{pmatrix}. \qquad (4)$$

The non-Hermitian parameters are $i\gamma_1$ and $\gamma_2$. Equation (4) has two order-2 EPs at $\phi_x = \frac{1}{2}\left(-\gamma_2 \pm \sqrt{\gamma_1^2 + \gamma_2^2}\right)$, $\phi_y = 0$, and they are connected by a branch cut, as shown in Figs. 2(b, c). When both EPs are encircled by a counterclockwise parametric loop, the Berry phase computed from Eq. (3) is $\Theta = \pi$. (When $\gamma_1 = \gamma_2 = 0$, the system is Hermitian and has a Dirac cone at $(\phi_x, \phi_y) = (0,0)$, which is a monopole in Berry curvature that gives rise to a Berry phase of $\pi$.) If only one of the EPs is encircled, the loop must cut through the branch cut such that two complete cycles are needed to produce a holonomy. From Eq. (3), the net Berry phase accumulated over two cycles is $\pi$.

To experimentally obtain $\Theta$, we need to obtain the REVs and LEVs under a parallel-transport gauge. In previous works concerning the measurement of the non-Hermitian Berry phase, only the REVs are obtained from experimental measurements. Owing to the transpose-symmetric property of the Hamiltonian, the LEVs were deduced by taking the transpose of the experimental REVs [15,28]. However, for the model in Eq. (4), $\widehat{H}_1 \neq \widehat{H}_1^\dagger$, $\widehat{H}_1 \neq \widehat{H}_1^*$, $\widehat{H}_1 \neq \widehat{H}_1^T$, such that the LEVs cannot be deduced from the REVs. They have to be separately measured.

We use two coupled acoustic resonant cavities to realize that model, as shown in Fig. 2(a). The system consists of two cuboid cavities ($2 \times 2 \times 12$ cm$^3$) that are coupled via a small tube. The resonant frequency of a single cavity is $\omega_0 = 9016$ rad/s, with a dissipative rate $\gamma_0 = -41.2$ rad/s. The coupling is tuned by adjusting the position of the coupling tube, which is 5.64 cm in length and 0.52 cm$^2$ in cross-sectional area. By varying the tube position (denoted as $D_z$ relative to the second cavity), the parameter $\phi_x$ in Eq. (4) changes from $-92.3$ rad/s to 92.3 rad/s. The detuning of the resonant frequency in the second cavity, which realizes $\phi_y$, can be adjusted within a tunable range of $-94.2$ to 94.2 rad/s. This adjustment is achieved by changing the volume of the cavity through the insertion of plasticine. The two non-Hermitian parameters of the system, $i\gamma_1$ and $\gamma_2$, are additional dissipation introduced in the first cavity and non-reciprocal coupling from cavity-1 to cavity-2, respectively. The additional loss is introduced by inserting sound-absorbing foam (at the bottom of cavity-1), and non-reciprocal coupling is realized by an active electroacoustic controller (AEC) [30,31]. Specifically, the AEC comprises four components: a microphone (Panasonic WM-G10DT502), a custom-printed circuit board (PCB) incorporating both a preamplifier and a phase shifter (Texas Instruments NE5532P), a power amplifier (Texas Instruments LM386), and a small loudspeaker. The microphone is placed at the top of the cavity to monitor the sound pressure (including amplitude and phase) in real-time. The measured signal is amplified by the preamplifier and sent to the phase shifter. Then, the modulated signal is amplified by the power amplifier to drive the loudspeaker at the base of the adjacent cavity. The phase shifter and amplifier are adjustable via potentiometers, allowing precise control over the phase and amplitude of the non-reciprocal coupling introduced into the system. The parameters in our experiment are $\gamma_1 = -19.7$ rad/s and $\gamma_2 = -40.8$ rad/s.

By tuning the two parameters $(\phi_x, \phi_y)$, we trace the two loops depicted in Figs. 2(b, c). At each parametric point, the steady-state responses are measured at both cavities, from which the REVs and LEVs are retrieved using the method mentioned above. The parallel transport gauge is enforced by compensating any difference in phase factors in front of the REVs and LEVs at neighboring steps [28]. To visualize the measured LEVs and REVs, we project them on the reference vector $|\text{Ref}\rangle = (1\ 0)^T$ and plot the results in Figs. 2(d, e). For the path enclosing both EPs (one EP), the arguments of the LEVs and REVs are reversed after one cycle (two cycles) of encirclement, implying a complete holonomy with a Berry phase of $\pi$. We further confirm the Berry phases by plotting their accumulations in Figs. 2(f, g). The excellent agreement of these experimental results with theoretical calculations validates our method for obtaining LEVs from Green's functions.

**3.2 the LEV of a non-Hermitian extended topological mode.**

In the second experiment, we apply the Green's function method to obtain the LEV of a topological mode in a 1D non-Hermitian SSH model, as shown in Fig. 3(a). The open-boundary Hamiltonian is

$$\widehat{H}_2 = \sum_{m=1}^{M-1} [v a_m^\dagger b_m + w a_{m+1}^\dagger b_m + (v + \delta) a_m b_m^\dagger + w a_{m+1} b_m^\dagger], \tag{5}$$

where $a_m^\dagger$ and $a_m$ ($b_m^\dagger$ and $b_m$) are the creation and annihilation operators of site-A (B) in $m$-th unit cell and $M = 6$ is the number of unit cells. The hopping terms are $v = -76.0$ rad/s, $w = -149.8$ rad/s, and $\delta$ denotes the non-reciprocal hopping, which is the only non-Hermitian parameter in the system. The corresponding acoustic model is shown in Fig. 3(b). The cavities are of the same shape and size as the two-level acoustic systems presented in Section 3.1. They are coupled with tubes with periodically alternating cross-sectional areas of 0.4 cm² and 0.8 cm². The specific values of the above parameters are obtained by fitting the responses using the Green's function.

When $|v| < |w|$ and $\delta = w - v$, the energy spectrum of the system under periodic boundary condition (PBC) traces two kissing loops in the complex plane, enveloping the purely real energy spectrum under open boundary condition (OBC), as shown in Fig. 3(c). At this time, the characteristic equations corresponding to the LEV and REV of the topological zero mode (TZM) can be described as:

$$\Psi_{m,A}^L = c_L \left(-\frac{v}{w}\right)^m, \Psi_{m,A}^R = c_R \left(-\frac{v+\delta}{w}\right)^m, \qquad \Psi_{m,B}^L = \Psi_{m,B}^R = 0, \tag{6}$$

where $\Psi_{m,A(B)}^{L(R)}$ are the entries of sublattice A (B) of the $m$-th unit cell in the LEV (REV) of the TZM, here $c_L$ and $c_R$ are normalization constants of LEV and REV, respectively. This configuration has an extended TZM [23,30], manifesting as a fully delocalized REV in the lattice. However, it is easy to see that the LEV of this state remains localized on the left side of the system. The theoretically computed REV and LEV are plotted in Figs. 3(d, e) as solid lines.

Such contrast in the profiles of the REV and LEV is ideal for testing the effectiveness of our Green's function-retrieval method. Experiments were performed by measuring each cavity's responses at the TZM's eigenfrequency ($\omega_0 = 9016 \text{ rad/s}$) when the source (loudspeaker) excites at each cavity individually. The measured REV and LEV are presented in Figs. 3(d, e) as markers. Excellent agreement with theory is seen, confirming the effectiveness of our method.

## 4. Conclusion

We present a reliable, easy-to-implement approach for obtaining REVs and LEVs from the measured responses of non-Hermitian systems and experimentally validate its effectiveness in acoustic experiments. The results also demystify the physical significance of LEVs – they are not merely mathematical objects, but physical entities with clear and unique observable effects in non-Hermitian systems. As such, our work finds an important missing piece in non-Hermitian physics and lays the foundation for further investigations of LEV-related phenomena in non-Hermitian physics.

**Acknowledgements**

G. M. thanks Zhao-Qing Zhang and Henning Schomerus for discussions. X. W. thanks Junjie Lu for helpful discussions. This work was supported by National Key R&D Program (2022YFA1404400), the Hong Kong Research Grants Council (RFS2223-2S01, 12301822), and the Hong Kong Baptist University (RC-RSRG/23-24/SCI/01, RC-SFCRG/23-24/R2/SCI/12).

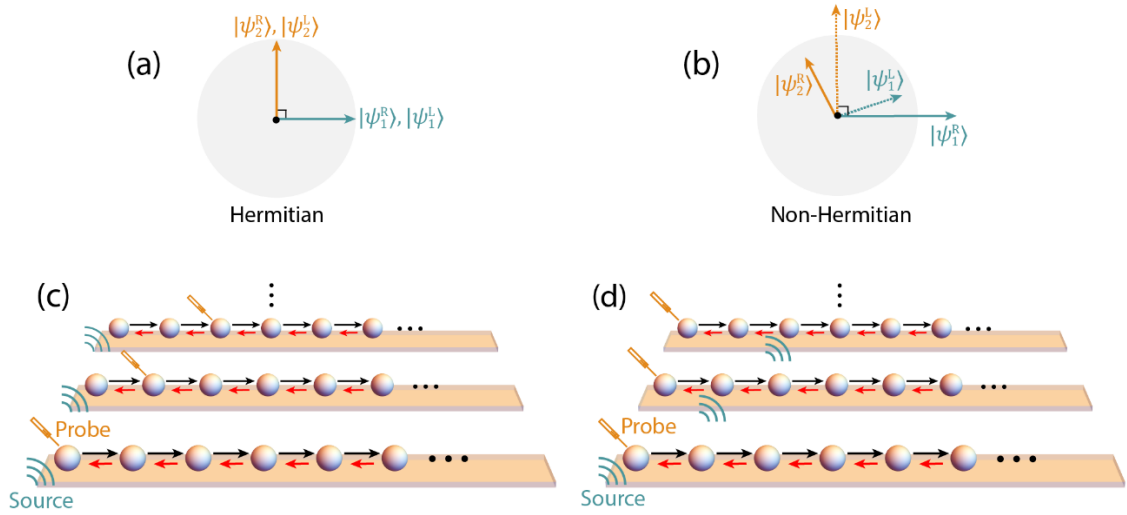

Fig. 1 (a) The relationship between LEVs and REVs in a Hermitian two-level system. (b) Non-Hermitian LEVs and REVs do not coincide. Here, the circle has a unity radius. (c) The REVs can be obtained from the steady-state response by fixing the excitation position and varying the measurement positions. (d) The LEVs are retrievable by fixing the measurement position and changing the excitation positions.

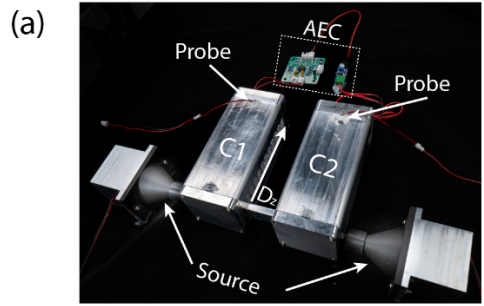

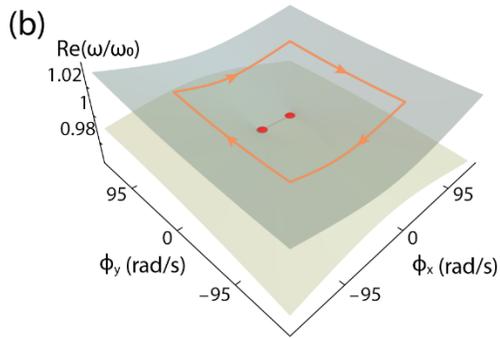 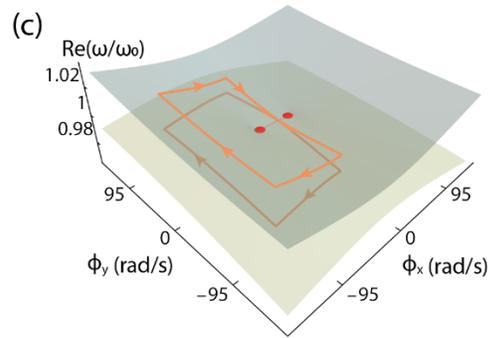

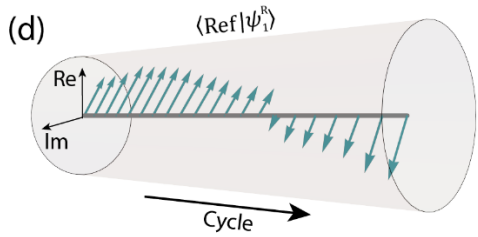 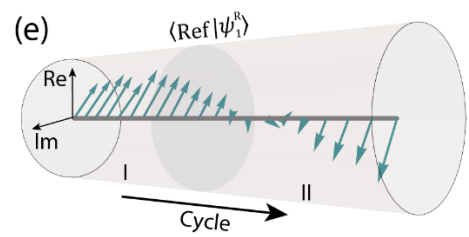

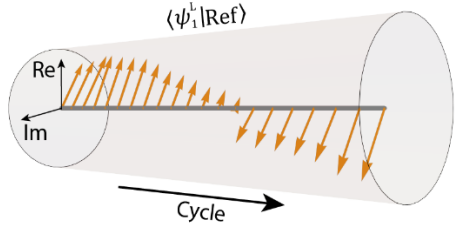 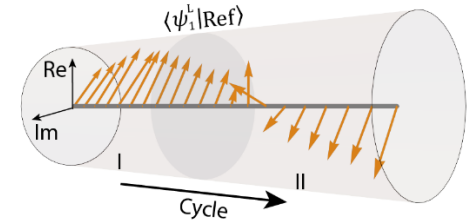

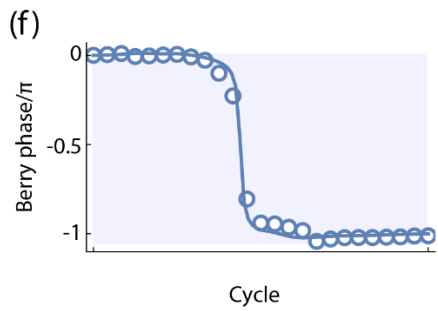 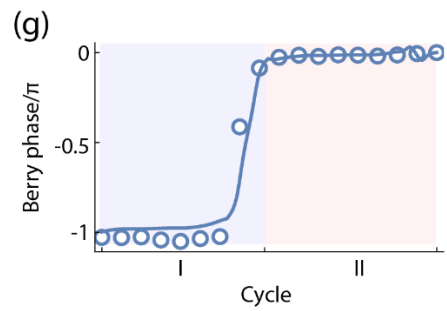

Fig. 2 (a) Coupled acoustic cavities realizing the two-level non-Hermitian model [Eq. (4)]. The sources (loudspeakers) and probes (microphones) are positioned at the bottom and top of two cavities, respectively. An AEC module facilitates non-reciprocal coupling between the two cavities. (b) and (c) A pair of EPs appearing on the $(\phi_x, \phi_y)$ parameter spaces, and solid lines with arrows represent parametric paths encircling the EP(s). In (b) the path encircles both EPs in a clockwise direction. In (c), the path encloses one EP. In this case, two complete cycles are needed to form a holonomy. (d) and (e) show the projection of the measured LEV $|\psi_1^L\rangle$ and REV $|\psi_1^R\rangle$ on the reference vector $|\text{Ref}\rangle = (1\ 0)^{\text{T}}$ under parallel transport. (f) and (g) show the cumulative Berry phases obtained from the REV and LEV retrieved from steady-state measurements.

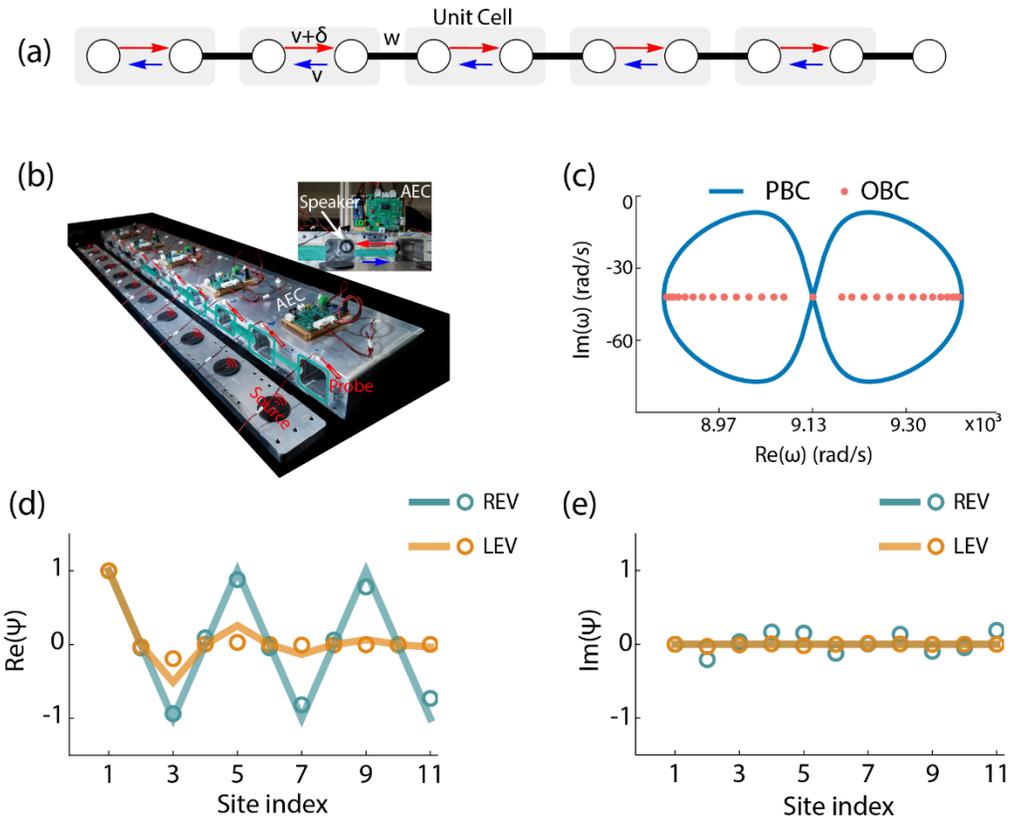

Fig. 3 (a) A schematic of the 1D non-Hermitian SSH model with an extended TZM [Eq. (5)]. (b) The acoustic cavity lattice realizing the model. Each acoustic cavity is equipped with a source and a probe. (c) The PBC and OBC spectra in the complex-$\omega$ plane. (d) and (e) display the measured real and imaginary components of REV and LEV associated with the TZM and $\delta = w - v = -73.8 \, \text{rad/s}$. The markers represent experimental results, and the solid lines are from theoretical calculations.

13